\documentclass{aastex62}
\begin{document}
\title{Tidal disruption near black holes and their mimickers} 
\correspondingauthor{Tapobrata Sarkar}
\email{tapo@iitk.ac.in}
\author{Pritam Banerjee}
\affiliation{Department of Physics\\ Indian Institute of Technology \\
Kanpur 208016, India}

\author{Suvankar Paul}
\affiliation{Department of Physics\\ Indian Institute of Technology \\
Kanpur 208016, India}

\author{Rajibul Shaikh}
\affiliation{Department of Physics\\ Indian Institute of Technology \\
Kanpur 208016, India}

\author{Tapobrata Sarkar}
\affiliation{Department of Physics\\ Indian Institute of Technology \\
Kanpur 208016, India}

\begin{abstract}
Black holes and wormholes are solutions of Einstein's field equations, both of which, from afar, look like a
central mass. We show here that although at large distances both behave like Newtonian
objects, close to the event horizon or to the throat, black holes and wormholes have
different tidal effects on stars, due to their respective geometries. We quantify this difference
by a numerical procedure in the Schwarzschild black hole and the exponential wormhole backgrounds,
and compare the peak fallback rates of tidal debris in these geometries.  
The tidal disruption rates in these backgrounds are also computed. It is shown that these quantities are 
a few times higher for wormholes, compared to the black hole cases.

\end{abstract}
\section{Introduction}

Tidal forces can tear apart a massive object due to the gravitational influence of another. 
This fact assumes importance in the context of supermassive black holes (BHs) with masses 
$M \sim 10^6~-~10^{10}M_{\odot}$ 
that are believed to exist at the center of most galaxies (see, e.g the reviews of \cite{Genzel}, \cite{Kormandy}, \cite{Alexander}).
Stars that come in the vicinity of such large masses are often disrupted by extreme gravity (\cite{LattimerSchramm}, 
\cite{EvansKochanek}, \cite{Syer}, \cite{MagorrianTremaine}, \cite{RRH}, \cite{Piran1}, see also the more recent review
by \cite{StoneRev}), when the tidal forces become comparable to, or larger than, the self gravity of the star. Such
tidal disruption events (TDEs) of stars in the background of a massive BH often produce an observable luminous flare, that
can reveal important properties of the stellar structure (\cite{Lodato1}, \cite{Lodato2}, \cite{Stone1}, \cite{Stone2}), 
as well as that of the BH (\cite{Kesden1}, \cite{ZLL}). Indeed, TDEs are known to be one of the
main physical processes responsible for the formation of accretion disks around BHs (\cite{Cannizzo}), a topic that has
received considerable attention of late, after the advent of the event horizon telescope (\cite{EHTAll}). Although several such TDEs have 
been observed, and seminal works have appeared in the literature over the last several decades, see, e.g \cite{FrankRees}, 
\cite{Lacy}, \cite{HillsNature}, \cite{ReesNature}, \cite{Phinney}, it is perhaps 
fair to say that a complete theoretical understanding of these processes is still lacking. This assumes importance 
in the light of the Large Synoptic Survey Telescope (LSST)\footnote{See https://www.lsst.org/} 
which is expected to provide more data on TDEs in the near future. 

In the context of a central mass $M$, the tidal radius $R_t \sim R_*\left(M/M_*\right)^{1/3}$, defined as the 
closest radial distance a stellar object of mass $M_*$ and radius $R_*$ 
can exist without getting tidally disrupted, is ubiquitous. This ``$1/3$ law'' is a standard
textbook result where one uses the fact that at the tidal disruption limit, the tidal force equals the force due to 
self gravity at the surface of a star. This Newtonian result is true for any central mass $M$, regardless of the
geometry that it produces, and it is perfectly legitimate to apply this to solutions of GR other than BHs, which might
not have any obvious interpretation as a singularity covered by an event horizon.
Indeed, while TDEs are believed to be common near galactic centers, relatively less attention has been paid in the literature
to the fact that such objects that do not have an event horizon (i.e are not black holes) can also tidally disrupt stars. One 
such example is provided by the wormhole (WH) geometry. Wormholes are exotic solutions of GR that connect two
different universes or two distant regions of the same universe via a throat, 
and continue to attract much attention, many decades after their inception. 
The original idea of the wormhole was related to a topology change of space-time by \cite{MW}, and 
several notable attempts to understand the formation of such wormholes as a result of a phase transition
were made in the early eighties by \cite{Sato1}, \cite{Sato2}, \cite{Sato3}. A traversable wormhole is one in which
material particles can tunnel through the throat, and \cite{FullerWheeler} 
demonstrated that the Schwarzschild wormhole (or the Einstein-Rosen bridge) is not traversable. 
A detailed construction of a traversable wormhole appeared in the work of \cite{MorrisThorne} and soon after, 
the construction of a ``time machine'' based on this idea appeared in the works of \cite{MTY}, and \cite{Novikov}. For an excellent 
exposition to related details, see the book by \cite{VisserBook} (see also the work of \cite{Lemos} for related literature). 
It is well known that typically the matter required to support wormhole geometries will 
violate standard energy conditions in GR (\cite{MorrisThorne}). Various attempts have however been made 
to evade such violations and as is well known by now, dynamical solutions (\cite{Sayan1}) or 
modified gravity (\cite{Lobo1}) might offer scenarios in which such a situation might be possible to envisage. 

Quite naturally, wormholes continue to be a popular theme ever since these works, both at the technical level and in 
popular imagination, and as such are often treated at the same level as horizonless astrophysical objects. Indeed,
a large amount of recent literature concerns astrophysical signatures of wormholes. 
For example, gravitational microlensing from wormholes were considered by \cite{Abe1}, 
\cite{Abe2},  while \cite{Taka} considered the cosmological constraints on these. Strong gravitational lensing by
wormholes were recently studied by \cite{tapo1} and \cite{tapo2}. Accretion disk properties of wormholes have
been studied in \cite{Harko1}, \cite{Harko2}, and the more recent review of \cite{Harko3}.

Part of the reason that wormholes are exciting is that these can effectively mimick black holes. This was
realized first by \cite{DamourSolodukhin}. These authors pointed out that several detectable features of black holes, 
like quasi-normal modes and accretion properties as well as theoretical features like no-hair theorems can be
very closely mimicked by wormhole solutions as well. In fact, after LIGO (\cite{LIGO}) announced the first 
evidences of gravitational wave detection, \cite{Cardoso1} reported that a class of wormholes that
have a thin shell of matter (with a specific equation of state) in the throat region can exhibit entirely similar
quasi-normal mode ringing at early times compared to black holes, with differences being clear only at late times. 
Subsequently, \cite{KZ} 
showed that specific wormhole solutions can in fact exhibit quasi-normal mode properties that are similar
to, or different from, black holes at all times. Clearly then, it is important and interesting to further the study of 
similarities between black holes and their wormhole mimickers, and in this work we study the differences 
in their TDEs. 

As we have mentioned, our understanding of TDEs is far from complete. 
The main reason for this is the complicated nature of TDEs which rule out an exact theoretical treatment, and one has
to resort to various approximations and costly numerical schemes. A part of the difficulty comes in due to the 
fact that one has to factor in effects of general relativity (GR) if stars come ``close enough'' to a BH (\cite{StoneRev}). 
To get an estimate of the numbers and the
approximations involved, let us consider a typical $10^6 M_{\odot}$ BH at the galactic center. The 
Schwarzschild radius for such a BH 
is about a solar radius, $R_{\odot}$. By a conservative estimate, we can assume that GR effects will be small 
beyond $\sim 10R_{\odot}$, in which region one can safely resort to a Newtonian approximation. 
Then, the innermost stable circular orbit of the BH being at $6R_{\odot}$, 
such effects will be important if the TDE occurs in this BH background, between $6R_{\odot}$ and $10R_{\odot}$. 
For more general parabolic or highly elliptical orbits, the periastron position in Schwarzschild black hole (SBH) backgrounds 
can be as small as $4R_{\odot}$, providing even further room for the effects of GR to come into the picture. Such effects
of GR on tidal disruption have been studied in several recent works in the context of BHs 
(see, e.g \cite{GReffects1}, \cite{Ishii-Kerr}, \cite{GReffects3}, \cite{GReffects4}, \cite{GReffects5}).
The question that we ask here is related to such effects in WH geometries, which are relatively less studied. 
Indeed, far from a BH or a WH, there is no way to distinguish them from their TDEs, as these are 
effectively Newtonian, and hence the tidal radius will follow the power law discussed earlier. However, near
the WH throat, something different is expected to happen. This is because near the throat of a WH, 
gravity is essentially repulsive in order to sustain the throat region, i.e to prevent it from collapsing. We therefore
expect that close to the throat, WHs should behave very differently from BHs. It is in this region that GR becomes
important, and hence one has to take care of GR effects carefully. 

In this paper, we point out that GR might modify the tidal radius exponent, so that a more
generic relation $R_t/R_* = \alpha \left(M/M_*\right)^\beta$, with $\beta \neq 1/3$ is a more appropriate working hypothesis 
in the region where GR effects are important. The quantity $R_*$ in this relation refers to the
radius of the star in the absence of strong gravity. What we find from a numerical analysis 
is that the index $\beta$, as well as the proportionality 
constant $\alpha$ have different values for black hole and wormhole geometries. While for BHs, $\beta < 1/3$, for
WHs on the other hand, $\beta > 1/3$. The index $\beta$ plays a crucial role in the dynamics of the tidally 
disrupted matter. As is known, after a TDE, roughly half of the stellar mass remains bound, and starts coming back
to the pericenter after a time $t_{min} \propto \left(\Delta \epsilon\right)^{-\frac{3}{2}}$, where $\Delta \epsilon$ is the 
dispersion in the energy of the tidal debris. Now, $\Delta\epsilon \sim R_t^{-2}$ (\cite{Phinney}), so clearly any change
in the index $\beta$ from its Newtonian value will be important in determining the accretion rate. In this paper, we perform
a numerical analysis adopting the method of \cite{Ishii-Kerr} and determine the index $\beta$. This is then used to 
determine the peak accretion rate, and we show how differences with the Newtonian limit arise for stars that
are tidally disrupted at distances close to the black hole event horizons or wormhole throats. 
Further, we study the rates of tidal disruptions in BH and WH backgrounds. 
These are also different due to reasons discussed above, and we quantify them in specific backgrounds. 

This paper is organized as follows. In the next section, we will briefly review the relevant 
formalism and some of the main consequences of the deviation from the Newtonian power law. In section 3, we
present our numerical analysis and determine this deviation, and revisit the consequences that it has on TDEs
in black hole and wormhole backgrounds. Section 4 ends this paper with a summary of the main results. 

\section{Methodology}

We will now discuss the methodology to be followed in this paper. These consist of three main ingredients, the
space-time metric, the FN frame, and the numerical method to compute the tidal disruption radius. 

\subsection{Space-time metrics} 

The generic rotating vacuum solution to Einstein's field equations is the Kerr black hole (KBH), represented by 
the axially-symmetric metric written in Boyer-Lindquist coordinates,
\begin{equation}
ds^2 = - \left( 1-\frac{2GMr}{c^2\Sigma} \right) c^2dt^2 -\frac{4 G M r a \sin^2 \theta}{c\Sigma} dt d\phi + \frac{\Sigma}{\Delta} dr^2 + 
\Sigma d\theta^2 + \left( r^2 + a^2 + \frac{2 G M r a^2 \sin^2 \theta}{c^2\Sigma} \right) \sin^2 \theta d\phi^2~,
\label{Kerr}
\end{equation}
where $ \Sigma = r^2 + a^2 \cos^2 \theta $ and $ \Delta = r^2 + a^2 -2 G M r/c^2 $. Here, $M$ is the ADM mass, and with $J$ being
the angular momentum of the KBH, $a = \frac{J}{Mc}$ 
represents the spin parameter such that $0\le a \le M$. The $a=0$ limit of the Kerr metric yields the static,
spherically symmetric Schwarzschild black hole (SBH) given by the metric
\begin{equation}
ds^2 = -\left(1-\frac{2GM}{c^2r}\right)c^2dt^2 + \frac{1}{1-\frac{2GM}{c^2r}}dr^2 + r^2d\theta^2 + r^2\sin^2\theta d\phi^2~.
\label{SBH}
\end{equation}
On the other hand, the quintessential example of a static traversable wormhole of the Morris-Thorne type is exemplified by
the metric 
\begin{equation}
ds^2=-e^{2\Phi(r)}c^2dt^2+\frac{1}{1-\frac{b(r)}{r}}dr^2+r^2(d\theta^2+\sin^2{\theta}d\phi^2)~.
\label{MTWH}
\end{equation} 
The example that we consider here is the exponential wormhole (EWH) given by the functions
\begin{equation} 
e^{2\Phi(r)} = e^{-\frac{2GM}{c^2r}},\;\;\;\; b(r) = \frac{2GM}{c^2}~,
\label{EWH}
\end{equation}
where $M$ is the ADM mass of the EWH. 

A useful comparison of TDEs in black hole and wormhole backgrounds can be made by considering the SBH and the 
EWH, as the position of the throat in the latter geometry is at $r_{th} = \frac{2GM}{c^2}$, which is the same location as the
Schwarzschild radius of the SBH. From a Newtonian perspective, an object that
is captured by a SBH will tunnel through the throat of an EWH of the same mass. 

\subsection{The Fermi-Normal frame}

We will consider parabolic (or highly elliptical) orbits in the backgrounds of the metrics mentioned above, in Fermi-Normal (FN) coordinates. 
The general formalism for space-times with spin is a convenient beginning point, from which static results can 
be obtained in the limit that the spin parameter goes to zero. For stationary backgrounds, we will restrict only
to the equatorial plane, where due to the axial symmetry, 
all the metric components are functions of $ r $ only. 
A star moving in a parabolic orbit close to the horizon of a BH or the throat of a WH 
will be influenced by the tidal fields of the BH and WH, respectively. GR effects are most conveniently studied 
by introducing a locally flat frame, called the Fermi Normal frame (\cite{Manasse-Misner}, \cite{marck}), that can move with 
the star along the geodesic where the time-like basis vector lies along the 4-velocity. The three other 
space-like vectors would be directed  perpendicular to the 4-velocity. 
This way it is possible to describe the inhomogeneous nature of gravity 
namely the tidal fields, in terms of the flat space-like coordinates. Exactly on the geodesic, the tidal force is zero. 
As one moves away perpendicular to the geodesic, the tidal fields increases.    

We will here display the FN coordinates for equatorial parabolic orbits in general spherically symmetric, static backgrounds, for simplicity, so that we can directly use them for both SBH and EWH. The same
can be constructed for such orbits in the Kerr geometry as well (\cite{marck}), and we will present numerical results on these later. 
On the equatorial plane, a general spherically symmetric, static metric can be written as
\begin{equation}
ds^2 = -A(r)dt^2 +B(r)dr^2+C(r)d\phi^2~.
\end{equation} 
Equatorial parabolic orbits specify the energy $ E $ and angular momentum $ L $ of a test particle as 
\begin{equation}
E = 1~,~~
L = \sqrt{\frac{C(1-A)}{A}}
\label{energyangmom}
\end{equation}

Now the Fermi Normal tetrad frame for an equatorial time-like geodesic can be written as
\begin{eqnarray}
\lambda_0 ^\mu &=& \{ \frac{E}{A},\frac{{\mathcal P}}{\sqrt{B}},0,\frac{L}{C} \} \nonumber\\
\lambda_1 ^\mu &=&\{\frac{\sqrt{B \text{C}} {\mathcal P}\cos\Psi}{\sqrt{B} \sqrt{A(C+L^2)}}-
\frac{EL\sin\Psi}{A \sqrt{C+L^2}},\frac{E\sqrt{A C}\cos\Psi}{A\sqrt{B(L^2+C)}}-\frac{L{\mathcal P}\sin\Psi}{\sqrt{B(L^2+C)}},0,-\frac{L\sqrt{L^2+C}\sin\Psi}{LC}\}\nonumber\\
\lambda_2 ^\mu &=& \{ 0,0,\frac{1}{\sqrt{C}},0 \}\nonumber\\
\lambda_3 ^\mu &=&\{\frac{\sqrt{B \text{C}} {\mathcal P}\sin\Psi}{\sqrt{B} 
\sqrt{A(C+L^2)}}+\frac{EL\cos\Psi}{A \sqrt{C+L^2}},\frac{E\sqrt{A C}\sin\Psi}{A\sqrt{B(L^2+C)}}
+\frac{L{\mathcal P}\cos\Psi}{\sqrt{B(L^2+C)}},0,\frac{L\sqrt{L^2+C}\cos\Psi}{LC}\}
\end{eqnarray}
where we have 
\begin{equation}
{\mathcal P} = \sqrt{\frac{E^2}{A}-\frac{L^2}{C}-1}~,~~\frac{d\Psi}{d\tau}= \frac{ELC'}{2\sqrt{ABC}(C+L^2)}
\end{equation}
The angle $ \Psi $ is introduced in order to parallel transport $ \lambda_1 ^\mu $ and $ \lambda_3 ^\mu $ along the time-like geodesic.

\subsection{Numerical Procedure}
\label{sec-3}
As shown by \cite{Ishii-Kerr}, the tidal potential in the FN frame can be expressed upto fourth
order in FN coordinates \{$x^0(=\tau),x^1,x^2,x^3$\} as,
\begin{eqnarray}
\phi_{\text{tidal}} &=&\frac{1}{2} C_{ij} x^i x^j + \frac{1}{6} C_{ijk} x^i x^j x^k + \frac{1}{24} \left[ C_{ijkl} + 
4 C_{\left(ij\right.} C_{\left.kl\right)} - 4 B_{\left(kl|n|\right.} B_{\left.ij\right)n} \right] x^i x^j x^k x^l + O(x^5)~,~
\label{eq.phi-tidal}
\end{eqnarray}
where the tensorial coefficients are defined as
\begin{equation}
C_{ij} = R_{0i0j}, ~~~ C_{ijk} = R_{0\left(i|0|j;k\right)}, ~~~ C_{ijkl} = R_{0\left(i|0|j;kl\right)}, ~~~ B_{ijk} = R_{k\left(ij\right)0}~.
\label{eq.Cij}
\end{equation}
Here, the Latin indices $i,j,k...$ run over the spatial components 1 to 3. The symbols `;' and `,' indicate the covariant derivative 
and the ordinary (partial) derivative respectively. Moreover, $  R_{0\left(i|m|j;kl\right)} $ is a summation over all possible 
permutations of the indices $i,j,k,l$ with $m$ fixed at its position, and then division by the total number of such permutations. 
For orbits where the stellar radius is comparable to the periastron distance, the third and fourth order corrections become important. 
We consider a fluid star in the FN frame. Since the frame provides a locally flat surroundings around the star, it can be described
by Newtonian gravity (\cite{Cheng-Evans2013}). The hydrodynamic equation for this fluid star in presence
of the tidal potential can then be expressed as
\begin{equation}
\rho \frac{\partial v_i}{\partial \tau} + \rho v^j \frac{\partial v_i}{\partial x^j} = - \frac{\partial P}{\partial x^i} - 
\rho \frac{\partial (\phi + \phi_{\text{tidal}})}{\partial x^i} +\rho \left[ v^j \left( \frac{\partial A_j}{\partial x^i} - 
\frac{\partial A_i}{\partial x^j} \right) - \frac{\partial A_i}{\partial \tau} \right] \label{eq.hydrodynamic}
\end{equation}
where the density, three-velocity ($ \frac{dx^i}{ d\tau} $) and pressure of the fluid are denoted by $\rho$, $v^i$ and $P$, 
respectively. The last term on the right hand side of Eq. (\ref{eq.hydrodynamic}) arises due to the gravito-magnetic force 
(\cite{gravito-magnetic}), and the corresponding vector potential reads, $A_i = \frac{2}{3} B_{ijk} x^i x^j$. 
Moreover, $\phi$ represents the Newtonian self-gravitational potential of the star, which obeys the usual Poisson equation,

\begin{equation}
\nabla^2 \phi = 4 \pi G \rho \label{eq.poisson}~.
\end{equation}
The co-rotational velocity field of the fluid star in the FN frame can be assumed as
\begin{equation}
v^i = \Omega [-\{x^3 - x_c \sin\Psi\},0,\{x^1 - x_c \cos\Psi\}]~, \label{eq.vel-field}
\end{equation}
where $ \Omega = d\Psi / d\tau $. In presence of the third order terms of the tidal potential and/or the 
gravito-magnetic terms, the correction term $x_c$ is non-zero, which corresponds to the fact that the rotational 
axis of the fluid star is different from the $x^2$-axis and the position of its center of mass slightly deviates from 
the origin of FN frame. Its magnitude is very small compared to the radius of the star. Now, to simplify our calculation 
by eliminating $\Psi$, we consider another frame in which the star is kept fixed so that the frame itself rotates with 
respect to the original FN frame. Coordinates of the new tilde frame are related to the FN coordinates by
\begin{equation}
\tilde{x}^1 = x^1 \cos \Psi + x^3 \sin \Psi, ~~~~ \tilde{x}^2 = x^2, ~~~~ \tilde{x}^3 = - x^1 \sin \Psi + x^3 \cos \Psi~. \label{eq.tilde-coord}
\end{equation}
Integrating the hydro-dynamic equation (Eq. (\ref{eq.hydrodynamic})) using the expression of $v^i$ (Eq. (\ref{eq.vel-field})), 
and then converting it to the tilde coordinates ($\tilde{x}^i$), we obtain
\begin{equation}
\frac{\Omega^2}{2} \left[ (\tilde{x}^1 - x_g)^2 + (\tilde{x}^3)^2 \right] = 
h + \phi + \phi_{\text{tidal}} + \phi_{\text{mag}} + C~,  \label{eq.basic-eq-2}
\end{equation}
where $C$ is an integration constant, $x_g=2x_c$, and $ h = \int \frac{dP}{\rho} $ . Here, 
$\phi_{\text{mag}}$ is the gravito-magnetic scalar potential arising from the last term on the right hand side of 
eq.(\ref{eq.hydrodynamic}) involving $A_i$. Eqs. (\ref{eq.poisson}) and (\ref{eq.basic-eq-2}) are the 
two equations that we solve numerically to obtain our results. Let us now consider the polytropic 
equation of state of the star as, $P=\kappa\rho^{1+1/n}$. 
Here, $\kappa$ is called the polytropic constant and $n$ is the polytropic index. The entire numerical 
calculation is performed in units, $c=G=M=1$. Moreover, we convert the basic equations (\ref{eq.poisson}) 
and (\ref{eq.basic-eq-2}) into dimensionless ones by using $\tilde{x}^i=p q^i$ to obtain numerical convergence, 
where the constant $p$ has dimension of length and $q^i$'s are dimensionless. Therefore, in terms of $q^i$'s, the basic equations become
\begin{equation}
\nabla_q^2 \bar{\phi} = 4 \pi \rho \label{eq.poissn-new}~,
\end{equation}
\begin{equation}
\frac{\Omega^2}{2} p^2 \left[ (q^1 - q_g)^2 + (q^3)^2 \right] = h(\rho) + p^2 \left(\bar{\phi} + 
\bar{\phi}_{\text{tidal}} + \bar{\phi}_{\text{mag}} \right) + C  \label{eq.basic-eq-2-new}~,
\end{equation}
where $ q_g = p^{-1} x_g $, $ \bar{\phi} = p^{-2}\phi $, $ \bar{\phi}_{\text{tidal}} = p^{-2}\phi_{\text{tidal}} $, 
$ \bar{\phi}_{\text{mag}} = p^{-2}\phi_{\text{mag}} $, and $ \nabla_q $ represents the Laplacian operator 
in $ q^i $ coordinates. Our purpose is to solve Eqs.(\ref{eq.poissn-new}) and (\ref{eq.basic-eq-2-new}) iteratively to
find the critical value ($ \rho_{\text{crit}} $) of the central density $\rho_c$ for which the fluid star just remains in 
stable configuration in presence of the tidal field. A star having central density ($\rho_c$) less than $ \rho_{\text{crit}} $ 
will be tidally disrupted. To obtain the numerical solution, we consider a three dimensional cubical grid system of 
equal grid size in every direction with $ 101 \times 101 \times 101 $ grid points. Complete detail of the step by 
step numerical procedure as well as determination of the constants, viz. $C$, $p$, $q_g$, etc from different boundary 
conditions can be found in \cite{kerr_tidal}.
\begin{table}[h!]
\begin{center}
\caption{Numerical values of $\xi_{R_*}$ and ${\mathcal I}$ for different $n$ \label{table1}}
\begin{tabular}{crrrr}
\tableline
\tableline
$n$ & $\xi_{R_*}$ & ${\mathcal I}~~~$ \\
\tableline
1 & $\pi~~~$ & 39.478 \\
3/2 & 3.654 & 34.106 \\
3 & 6.897 & 25.362 \\
\tableline
\end{tabular}
\end{center}
\end{table}
Once the critical central density ($ \rho_{\text{crit}} $) and the corresponding density profile of the star are obtained 
from the numerical analysis mentioned above, we find out the critical mass of the star by integrating the density profile. 
Specifically, introducing the dimensionless parameters $\Theta$ and $\xi$, defined as, $\rho = \rho_c \Theta^n$ 
and $r=\bar{r}~\xi$, where $\rho_c$ is the central density and $\bar{r}$ is a constant having dimension of length, 
the critical mass of the star comes out to be
\begin{equation}
M_\star=\frac{R_\star^3}{\xi_{R_\star}^3}\rho_c \left[4\pi \int_{0}^{\xi_{R_\star}}\Theta(\xi)^n ~ \xi^2 d\xi\right]
=\frac{R_\star^3}{\xi_{R_\star}^3}\rho_c \mathcal{I}~,
\label{mass}
\end{equation}
where in the last relation one has to use the critical value $\rho_{\rm crit}$ of the central density $\rho_c$.
\begin{figure}[h!]
\centering
\centerline{\includegraphics[scale=0.5]{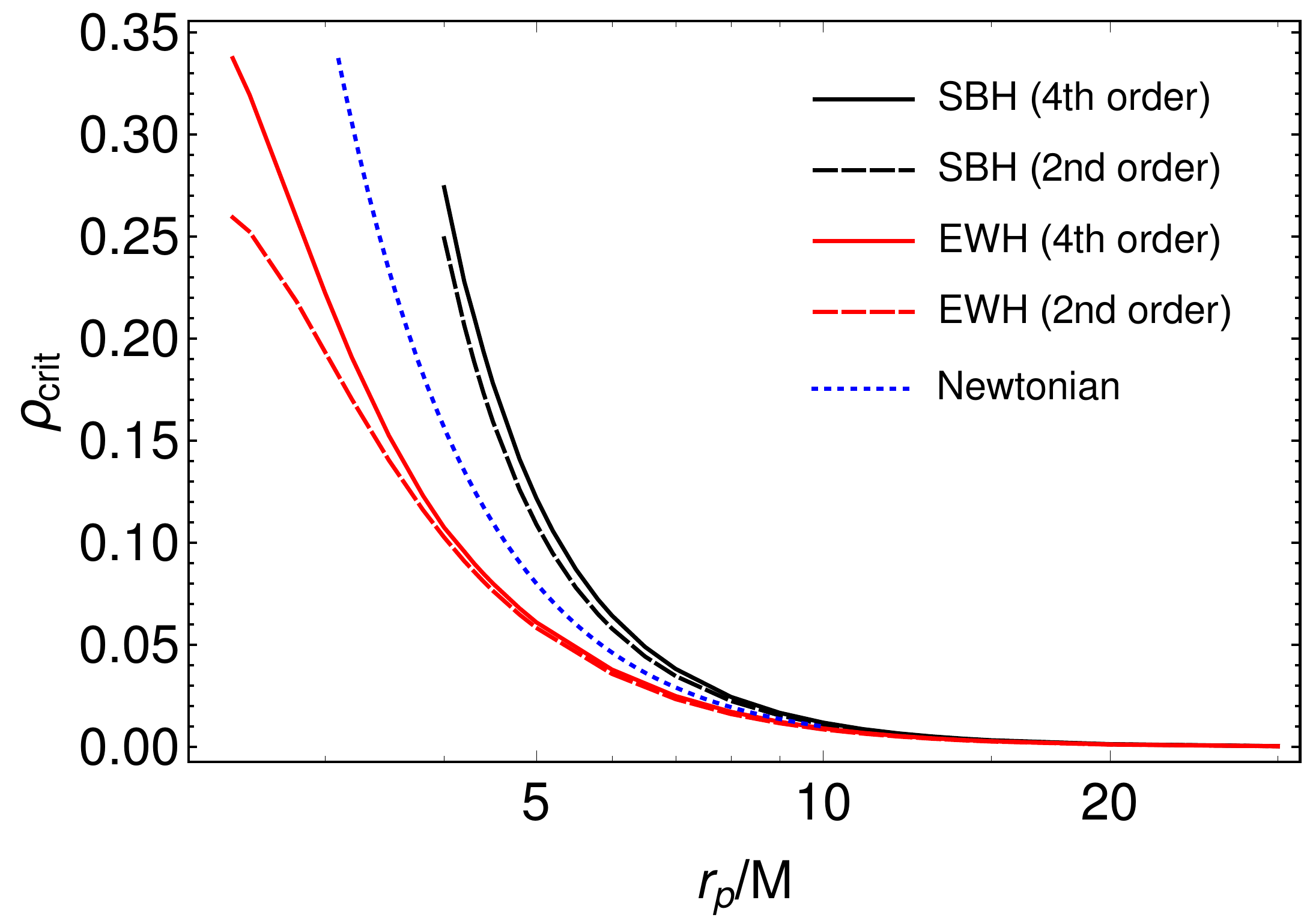}}
\label{fig.rho_WH_vs_BH_para}
\caption{$ \rho_{\text{crit}} $ as a function of the periastron position $ r_p $ in logarithmic scale for the EWH
and the SBH. Here, the masses of the EWH and the SBH have been set to $1$ and so is the radius of the star. 
The Newtonian power law behaviour is also shown.}
\label{fig1}
\end{figure}
We tabulate the values of $\xi_{R_\star}$ and $\mathcal{I}$ in table (\ref{table1}), for different values of $n$. Here, 
$\xi_{R_\star}$ corresponds to the value of the Lane-Emden parameter $\xi$ at the surface of the star. Once we calculate the critical mass of the star, we fit this to a relation of the form
\begin{equation}
\left(\frac{R_t}{R_\star}\right)=\alpha \left(\frac{M}{M_\star}\right)^{\beta} \label{eq:fitting_eqn}~.
\end{equation}
We emphasize that all terms in Eq. (\ref{eq:fitting_eqn}) are in $c=G=M=1$ units as described above. In Fig. (\ref{fig1}), 
we show the variation of the critical central density ($ \rho_{\text{crit}} $) as a function of the periastron position 
($r_p$) for both the EWH and SBH, where we have taken the masses of the EWH and SBH to be unity, and have 
also considered, $R_\star=1$. 
Table (\ref{table2}) shows the results of the fit mentioned in Eq. (\ref{eq:fitting_eqn}).
\begin{table}[h!]
\begin{center}
\caption{Numerical values of $\alpha (n)$ and $\beta (n)$ for different $n$, for various backgrounds and fitting ranges \label{table2}}
\begin{tabular}{crrrrrrrrrrr}
\tableline
\tableline
Object & Fitting Range & $\alpha (1)$ & $\beta (1)$ & $\alpha (1.5)$ & $\beta (1.5)$ & 
$\alpha(3)$ & $\beta(3)$ \\
\tableline
SBH (2nd) &$4\leq r/M\leq 12$ &2.752 & 0.303  &2.740 &0.303 &2.763  &0.305 \\
$~$ &$100\leq r/M\leq 2000$ & 2.335 &0.333  &2.317 &0.333 &2.369  &0.333 \\
EWH (2nd) & $2.4\leq r/M\leq 12$ &1.842 &0.374 &1.832 &0.375  &1.835  &0.380 \\
$~$ &$100\leq r/M\leq 2000$ &2.318 &0.333 &2.317 &0.333  &2.369  &0.333 \\
SBH (4th) &$4\leq r/M\leq 12$ &2.866 & 0.298  &2.852 &0.299 &2.862  &0.302 \\
$~$ &$100\leq r/M\leq 2000$ & 2.338 &0.333  &2.320 &0.333 &2.372  &0.333 \\
EWH (4th) & $2.4\leq r/M\leq 12$ &1.918 &0.370 &1.904 &0.371  &1.917  &0.374 \\
$~$ &$100\leq r/M\leq 2000$ &2.323 &0.333 &2.320 &0.333  &2.372  &0.333 \\
KBH (4th) (a=0.9)&$1.8\leq r/M\leq 12$ &2.634  &0.314  &2.627 &0.314  &2.664  &0.315 \\
$~$ &$100\leq r/M\leq 2000$ &2.336  &0.333  &2.338 &0.333  &2.373  &0.333 \\
KBH (4th) (a=0.99) &$1.8\leq r/M\leq 12$ &2.591 &0.317 &2.584 &0.318 &2.621  &0.319 \\
$~$ &$100\leq r/M\leq 2000$ & 2.336 &0.333 &2.338 &0.333 &2.373  &0.333 \\
\tableline
\end{tabular}
\end{center}
\end{table}

\section{Analysis}
\label{sec-3.1}

Here, we will present the implications of the discussion above. 
First, note that for a star to be tidally disrupted by a black hole (rather than being swallowed as a whole), we require
the tidal radius to be greater than the radius of the event horizon. In contrast, for a wormhole, if the tidal radius is less
than the radius of the wormhole throat, the object can tunnel through the throat into the universe on the other side. 
For the SBH, as already mentioned, the closest periastron position of the orbit is characterized by $r/M=4$. From 
our numerical analysis, we find that for a star to be tidally disrupted at this value of the radius, one has, with $M=1$,
$\rho_c = 0.249$. The tidal disruption radius for the EWH of the same mass for the same value of $\rho_c$ is 
$r/M = 2.531$. These values are conveniently tabulated in table (\ref{table3}), where we have also shown the value
of the tidal radius for which a star is disrupted at $R_t/M=4$ as given by a purely Newtonian analysis. The differences
in the values of the various columns of $R_t$ are then purely due to GR effects. The first two rows in table (\ref{table3}) 
corresponds to a second order GR calculation while in the last two rows we show the results up to fourth order
in the tidal potential, which is important only for systems like a SBH-neutron star binary.  
\begin{table}[h!]
\begin{center}
\caption{Numerical values of the tidal radius in units of $M=1$ at second and fourth order \label{table3}}
\begin{tabular}{crrrr}
\tableline
\tableline
Order & $\rho_c$ & $R_t$ (N) & $R_t$ (SBH) & $R_t$ (EWH) \\
\tableline
Second & 0.249 & 3.430 & 4.00 & 2.531 \\
$~$ & 0.157 & 4.00 & 4.532 & 3.350 \\
\hline\hline
Fourth & 0.274 & 3.320 & 4.00 & 2.740\\
$~$ & 0.157 & 4.00 & 4.700 & 3.460\\
\tableline
\end{tabular}
\end{center}
\end{table}
The value of $\rho_c$ for which tidal disruption occurs at $r/M=4$ for the SBH gives us the maximum possible
difference between effects of TDEs in SBH and EWH backgrounds. From table (\ref{table3}), we see that for 
a $10^8M_{\odot}$ SBH and an EWH of similar mass, the difference in the tidal disruption
radius of a solar mass star is $\sim 10^{-5}$ light years. In table (\ref{table4}), we show the typical
masses and radii of stars for which $\rho_c = 0.249$. For the last two rows in table (\ref{table4}), the fourth
order results listed in table (\ref{table2}) are required. 
\begin{table}[h!]
\begin{center}
\caption{Numerical values of the masses and radii of stars with $\rho_c=0.249$ \label{table4}}
\begin{tabular}{crrrr}
\tableline
\tableline
$M/M_{\odot}$ & $~~M_*/M_{\odot}$ & $R_*$ (cm) \\
\tableline
$10^8$ & $1~~~$ & $4.69 \times 10^{10}$ \\
$10^6$ & $0.7~~$ & $1.93 \times 10^9$ \\
$10^5$ & $0.7~~$ & $4.16 \times 10^8$ \\
$10$ & $1.4~~$ & $1.09 \times 10^6$ \\
$5$ & $1.4~$ & $6.89 \times 10^5$\\
\tableline
\end{tabular}
\end{center}
\end{table}

Now, once the star is tidally disrupted, the spread of the specific energy of the debris is given from the expression of
\cite{Stone3} as $\Delta\epsilon = GMR_*/R_t^2$.
Then, one obtains the minimum time required for the debris to come back to the
pericenter (\cite{Lodato2}) as 
\begin{equation}
t_m = \frac{2\pi GM}{\left(2\Delta\epsilon\right)^{3/2} }\propto R_t^{3}.
\label{tmin}
\end{equation}
The peak fallback rate, which is proportional to $M_*/t_m$ can then be compared to the Eddington accretion rate, 
and for $R_* \sim R_{\odot}$ and $M \sim 10^6 M_{\odot}$, is typically a few orders of magnitude higher than
the latter. Then, we see that the maximum value of the ratio 
\begin{equation}
\frac{t_{m,EWH}}{t_{m,SBH}} = \frac{R_{t,EWH}^3}{R_{t,SBH}^3} = \left(\frac{2.53}{4}\right)^3 = 0.253~.
\label{mainmod}
\end{equation}
Denoting the peak fallback rate as ${\dot M_p}$, the ratio of the rates, which is proportional to $1/t_m$ is then obtained as
${\dot M_{p,EWH}}/{\dot M_{p,SBH}}=3.953$

Finally, we compute the tidal disruption rate of a star by Schwarzschild
black hole or exponential wormhole. We assume a Maxwellian distribution of star
velocities with a given number density $ n_d $ and velocity dispersion $ \sigma $
far from the BH or WH so that they follow highly elliptic or parabolic orbits to
reach the minimum periastron position $ r_p $ near the BH or WH. In the process, stars are
either tidally disrupted or directly captured by the BH or the WH depending on the $ r_p
$. If $ r_p $ is less than the tidal radius $ R_t $ it can be tidally disrupted.
They are directly captured when the orbits enter the event horizon or the throat without tidal disruption. We
assume that a star with a specific orbital energy $ E $ and angular momentum $ L $
approaches a BH or a WH. Following \cite{Kesden1}, the capture rate is given by,
\begin{equation}
\Gamma_{\rm cap} = \int_{0}^{L_{\rm cap}}\frac{\partial\Gamma}{\partial L}dL = \frac{\sqrt{2
\pi} n_d L_{\rm cap}^2}{\sigma}
\end{equation}   
where, $ L_{\rm cap} $ is the specific orbital angular momentum of the star below which
it enters the event horizon or the throat without being tidally disrupted. 
\begin{figure}
\gridline{
\fig{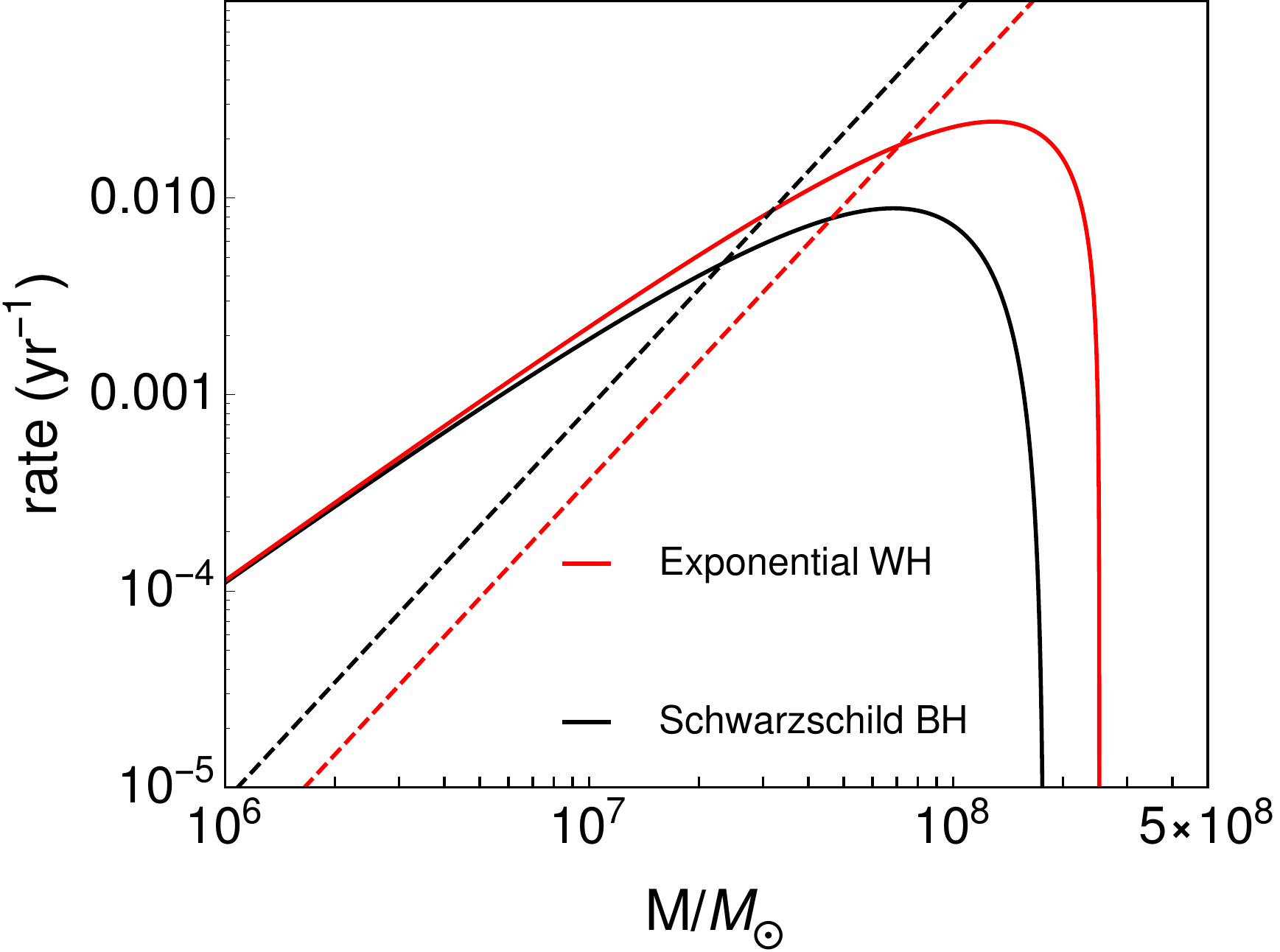}{0.5\textwidth}{(a)}
\fig{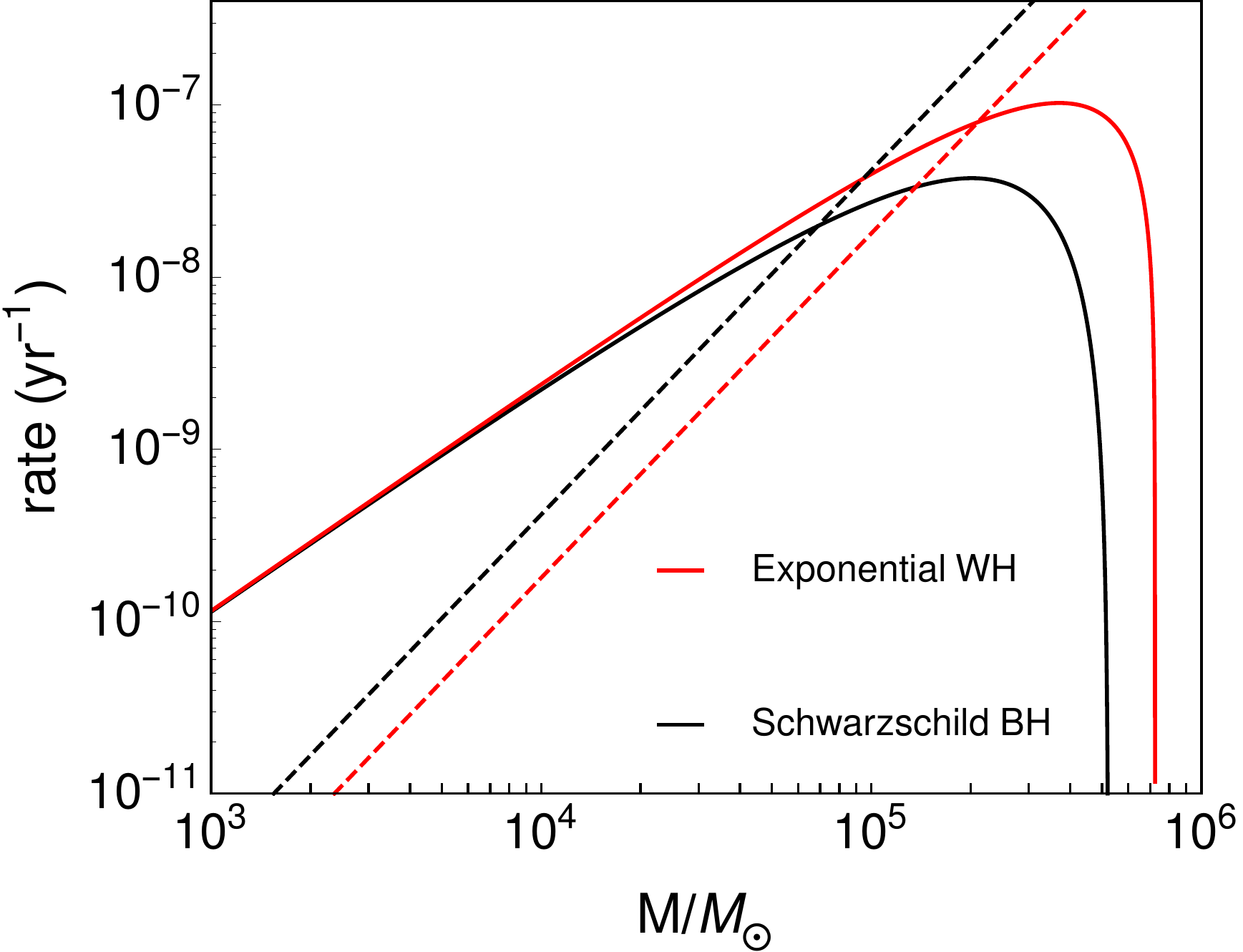}{0.5\textwidth}{(b)}}
\caption{Variation of $ \Gamma_{\rm TDE} $ with M (mass of the BH or WH) for a star having (a) mass $ M_{\odot} $ and radius $ R_{\odot} $, and (b) mass $ 0.7 M_{\odot} $ and radius $ 10^9~{\rm cm} $. The solid red curves represent EWH and the solid black curves correspond to SBH. The dashed lines denote the capture
rate $ \Gamma_{\rm cap} $, with the same color coding. }
\label{fig2}
\end{figure}

\begin{figure}[h!]
\centering
\centerline{\includegraphics[scale=0.5]{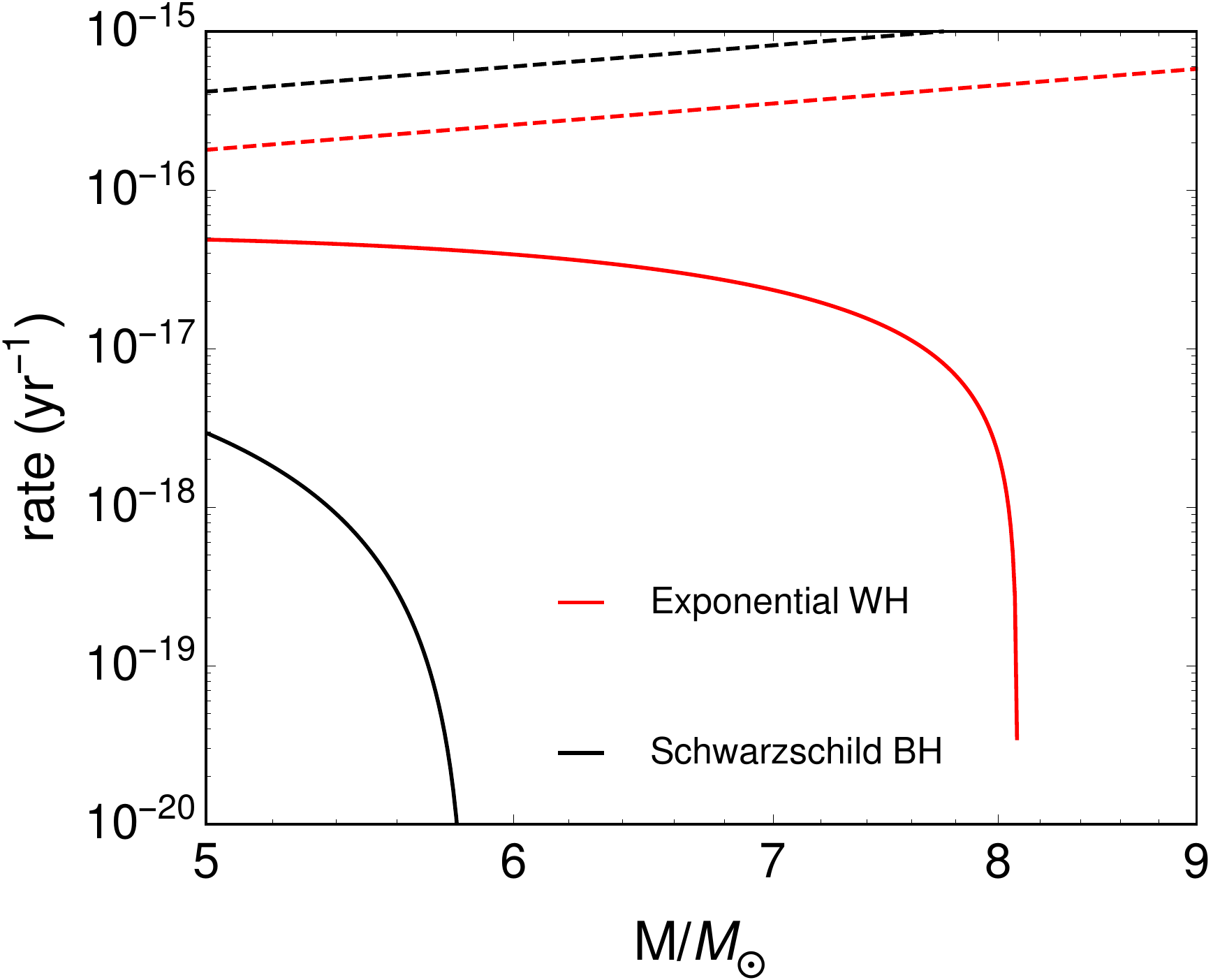}}
\caption{Variation of $ \Gamma_{\rm TDE} $ with M (mass of the BH or WH) for a star having mass $ 1.4 M_{\odot} $ and radius $ 10^7~{\rm cm} $. The color coding is the same as in Fig. (\ref{fig2}).}
\label{fig3}
\end{figure}
Similarly, tidal disruption rate $ \Gamma_{\rm tidal} $ is obtained by evaluating the
integration upto $ L_{\rm tidal} $, the specific orbital angular momentum below which
the star enters the tidal radius $ R_t $. Now, the tidal disruption event rate is
obtained as, $ \Gamma_{\rm TDE}=\Gamma_{\rm tidal}-\Gamma_{\rm cap} $. 
In order to find $ L_{\rm cap} $ or $ L_{\rm tidal} $ we first choose the mass and radius of a
star in units of $ c=G=M=1 $, where $ M $ is the BH or WH mass. Then we put these values in eq.(\ref{mass})
to find out the critical central density for a given value of $n$. Once we have the critical central density,
we can calculate the minimum possible $ r_p $ which the star can
attain without getting tidally disrupted, i.e., $ r_p=R_t $. Now, we put this $ r_p $ in
eq.(\ref{energyangmom}) to calculate $ L_{\rm tidal} $, whereas $ L_{\rm cap} $ is the value of $L$ for which 
$ r_p = 4M$ for the SBH and $2M$ for the EWH. 

Considering a constant density $ n_d=10^5 {\rm pc}^{-3} $ and $ \sigma=10^5{\rm cm/s} $, the
dependence of TDE rate with $ M $ is shown in figures (\ref{fig2}(a)), (\ref{fig2}(b)) and (\ref{fig3}). 
Variation of the tidal disruption event rate $ \Gamma_{TDE} $ for solar
mass stars, white dwarfs and neutron stars with BH or WH mass is shown. Here, the
solar mass star with mass $ M_{\odot} $ and radius $ R_{\odot} $, white dwarf
with mass $ 0.7 M_\odot $ and radius $ 10^9 {\rm cm} $, and neutron star with mass $ 1.4
M_\odot $ and radius $ 10^5 {\rm cm} $ are considered. The polytropic index is fixed at $
n=1 $, along with $ n_d = 10^5 {\rm pc}^{-3} $ and $ \sigma = 10^7 {\rm cm/s} $. The solid red lines 
correspond to the EWH and the solid black lines for the SBH. The dashed lines denote capture
rate $ \Gamma_{\rm cap} $, with the same color coding.
For main sequence stars, the
TDE rate vanishes for $ M \sim 10^8 M_\odot $ or more. This suggests that main sequence stars
are directly captured before getting tidally disrupted. White dwarfs are tidally
disrupted if $ M \sim 10^5 $ or less, and neutron stars are only disrupted by
stellar mass BH or WHs. The capture rate for the EWH is less than that
of the SBH. This is due to the fact that the capture radius ($ 2M $) for
exponential WH is smaller than the Schwarzschild BH. This suggests that the background
geometry can make significant role in TDE rates.

\section{Summary and Discussion}
\label{sec-5}

Wormholes have various observational features that are similar to black holes and hence these
can act as black hole mimickers. In this paper we have studied to what extent these behave
differently as far as tidal disruptions are considered. This is important and interesting, as such
tidal disruptions are one of the main reasons for the formation of accretion disks. From a 
Newtonian perspective, the ubiquitous ``1/3 law'' mentioned in the introduction 
determines the tidal radius for a given 
central mass, irrespective of its geometry. We have shown here that this rule is modified by
strong gravity, in a way that can serve as a distinction between BHs and WHs. The physical reason for
such a modification is that the physics near the black hole event horizon and that near the
wormhole throat are very different. In the latter case, to sustain the throat so that it does not
collapse, gravity might behave differently from that near a black hole horizon. Indeed, the
deviation from the Newtonian result for the tidal radius is seen to be of opposite signs in
the two backgrounds. 

These effects might be important for observational distinctions, for example from data expected from the LSST. 
We have considered two such effects here. Firstly, we find that the peak fallback
rate for wormholes might be about $4$ times higher than that of a black hole with similar mass.
Secondly, the tidal disruption rates for the wormhole examples we consider are greater than
those for the corresponding black holes. 
Our analysis here is numerical, and specialized to a static scenario. A time-dependent 
analysis of tidal disruptions will be a natural extension of this work.


\end{document}